\documentclass[twocolumn]{aastex631}
\usepackage{booktabs}

\begin{document}

\title{\textbf{Quantifying the Distribution of Europium in the  Milky Way Disk: Its Connection to Planetary Habitability and the Source of the r-Process}}

\author[0009-0005-5223-1606]{Evan M. Carrasco}
\affiliation{Department of Astronomy and Astrophysics, University of California, Santa Cruz, CA 95064, USA}

\author[0000-0003-0509-2656]{Matthew Shetrone}
\affiliation{University of California Observatories, CA 95064, USA}

\author[0000-0003-3573-5915]{Francis Nimmo}
\affiliation{Department of Earth and Planetary Sciences, University of California, Santa Cruz, CA 95064, USA}

\author[0000-0003-2558-3102]{Enrico Ramirez-Ruiz}
\affiliation{Department of Astronomy and Astrophysics, University of California, Santa Cruz, CA 95064, USA}

\author[0000-0001-5091-5098]{Joel Primack}
\affiliation{Physics Department, University of California Santa Cruz, 1156 High Street, Santa Cruz, CA 95064 USA}

\author[0000-0002-7030-9519]{Natalie M. Batalha}
\affiliation{Department of Astronomy and Astrophysics, University of California, Santa Cruz, CA 95064, USA}

\author[0009-0007-1664-4774]{Brady Lobmeyer}
\affiliation{Department of Astronomy and Astrophysics, University of California, Santa Cruz, CA 95064, USA}

\begin{abstract}

The astrophysical site of the r-process remains one of the most pressing questions in stellar nuclear synthesis. Although multiple theoretical sites have been proposed, with some observational counterparts available, the current Galactic distribution cannot be reproduced from a single consistent injection site. To disentangle the prospective sites of r-process production, the distribution of r-process elements in the Galaxy today must be scrutinized. In this study, we find that the intrinsic star-to-star distribution of the r-process element europium (Eu) at a fixed metallicity and temperature has a small intrinsic [Eu/H] scatter of 0.025 dex. In addition to a small dispersion, we demonstrate an anticorrelation between [Eu/$\alpha$] and [$\alpha$/H] consistent with either r-process production being metallicity dependent in core-collapse supernova and/or being produced in double neutron star mergers with a delay time distribution greater than $t^{-1}$. Furthermore, using Eu as a proxy for the radioactive r-process elements U and Th, and assuming that rocky planetary abundances reflect their parent star's composition, we show how these elements play a key role in the evolution of the magnetic dynamo on Earth-mass planets. Specifically, we find that only above [$\alpha$/H] $>-0.25$ do most stars' planetary systems meet the threshold abundance of [Eu/$\alpha$] $<+0.06$ to support a persistent magnetic dynamo, supporting the notion of a ``habitable metallicity range'' in the Galactic disk.

\end{abstract}

\section{Introduction} \label{sec:intro}
The heaviest elements on the periodic table are formed by neutron capture. This occurs in two flavors: the slow ``s-process" and rapid ``r-process". Although the s-process is relatively well understood to take place in Asymptotic-Giant-Branch (AGB) stars, the exact source of the r-process has remained elusive, primarily owing to the extreme neutron-rich environment required to sustain the r-process. Multiple theoretical sites such as exotic core collapse supernovae (CC-Sne) (e.g. magnetars \citealt{2018ApJ...864..171M,2024ApJ...969..141Z}; collapsars \citealt{2024PhRvD.110h3024D,2024arXiv241002852I}), and the ejecta of binary neutron star-merger events (NS-NS mergers, \citealt{2018IJMPD..2742005H}) have been proposed, although the dominant production mechanism is still debated \citep{Cescutti2005,Macias2019,Holmbeck2023}. NS-NS mergers have been shown to be a robust channel of r-process synthesis in the observational follow-up of the kilonova counterpart SSS17a/AT2017gfo \citep{2017Sci...358.1556C} to the gravitational wave event GW170817 \citep{2017PhRvL.119p1101A}. However, observed yields cannot reproduce the current Galactic distribution of r-process elements, suggesting multiple possible sites of production or changing conditions of synthesis throughout Galactic history. For an overview of the current discussion and outstanding issues, see \cite{2019EPJA...55..203S,Chen2024,2024arXiv240608630M}. Several studies have measured the stellar abundances of r-process elements in F-, G-, and K-type stars in the Milky Way Disk (e.g., \citealt{Prochaska_2000,2001A&A...376..232M,2005A&A...433..185B,2006MNRAS.367.1329R,Bensby2014,Battistini2016,DelgadoMena2017,2018A&A...619A.143G,Botelho2019,2022MNRAS.516.3786M}). To break the degeneracy of the r-process sites, the current Galactic distribution must be examined to understand the yields of individual events and the time delay of material injection in conjunction with Galactic chemical simulations to rule out competing theories. The importance of quantifying the Galactic r-process distribution is twofold: First, the distribution of r-process elements reflects the sites of origin. Secondly, the thermal evolution of rocky planets depends on the abundance of r-process elements to drive radiogenic heating.

\begin{table*}[ht!]
\centering
\begin{tabular}{ l|c|c|c|c|c}
    \toprule
    Sample & Zero-point & [$\alpha$/H] Slope & $T_{\rm eff}$ Slope & SD & CM SD \\
    \midrule
    \citet{DelgadoMena2017} & 0.521 $\pm$ 0.055  & 0.844 $\pm$ 0.016 & -9.14E5 $\pm$ 9.669E-6 & 0.071 & 0.057 \\ 
    \citet{Bensby2014} \& \citet{Battistini2016} &  -0.527 $\pm$ 0.099 & 0.768 $\pm$ 0.022 & 9.78E-5 $\pm$ 1.71E-5 & 0.081 & 0.065 \\
    \bottomrule
\end{tabular}  
\caption{Second order fitting polynomials for the DM and BB samples, fit to temperature ($T_{\rm eff}$)(K) and metallicity ([$\alpha$/H])(dex). A function of detrended [Eu/H] is created by subtracting the above expressions from [Eu/H] for each sample. The cross-matched expression is given as an average of the DM and BB expressions. Additionally the standard deviation (SD) of detrended [Eu/H] (dex), and the standard deviation of the cross matched (CM) detrended [Eu/H] (dex) is given. Coefficients A, B, and C in Equation \ref{eq:plane} correspond to the $T_{\rm eff}$ Slope, [$\alpha$/H] Slope and Zero-point columns respectively.}
\label{tab:Contour}
\end{table*}

While the astrophysical site of the r-process remains uncertain, the heavy elements produced are crucial ingredients in driving the thermal evolution of rocky planets. Long-lived r-process radioisotopes thorium-232 ($^{232}$Th half-life 14.05 Gyr), uranium-235 ($^{235}$U half-life 703 Myr), and uranium-238 ($^{238}$U half-life 4.47 Gyr) are embedded in the mantle/crusts of rocky planets.\footnote{Potassium-40 ($^{40}$K) is another significant contributor included in our later modeling; however, because of its relatively short half-life of 1.25 Gyr its contribution is modest at late times \citep[see for further ][]{Nimmo2020}.} Over geological times scales (Gyrs) these elements undergo decay, releasing energy into the mantle which drives the geodynamic process important to the evolution of surface conditions (e.g.,  \citealt{Frank2014,Jellinek2015,Botelho2019,Unterborn2015,Foley2018,Quick2020,Zhang2022,Boujibar2020,Wang2020,Mello2023}). Here we focus on the generation of a magnetic field and its effect on atmospheric preservation. The magnetic dynamo of an Earth-like planet is driven by convection in its liquid core, whose strength is moderated by thermal exchange across the core-mantle boundary \citep{Labrosse2014}. This rate of exchange and, by extension, the vigor of the dynamo are sensitive to radiogenic heating. Excessive mantle heating (caused by an overabundance of radioactive elements) reduces the heat extracted from the core and thus the thermal buoyancy available to drive core convection. This ultimately leads to periods of dynamo failure \footnote{A similar effect may occur at Io, driven by excessive tidal heating in the mantle \citep{Wienbruch1995}.} \citep{Nimmo2020,Labrosse2015,Boujibar2020}. A robust magnetic dynamo is considered a defining feature of Earth's habitable biosphere, protecting the surface from intense solar irradiation, preserving the atmosphere and preserving liquid-water inventories. The terrestrial geodynamo has persisted for at least the last 3.7~Gyr with no interruptions \citep[e.g.,][]{Nichols2024}, although the magnetic field has occasionally dropped to as low as one-tenth of its mean value \citep{Bono-etal:2019}.

If we wish to understand how the geodynamos of rocky planets similar to Earth have evolved as a function of radiogenic heating, we must first understand how these radioactive elements are distributed in nearby stars. Although the mantle abundances of exoplanets are inaccessible directly by current techniques, given that the planet and star form from a common natal molecular cloud, the planetary abundances of refractory elements (such as the isotopes of interest) should reflect the abundances of the host star \citep[e.g.,][]{Naiman2018}. Granted, lighter elements have complicated radial dependencies due to ice lines, here we assume that refractory metals and radioactive elements are distributed homogeneously between star and rocky planet. U and Th (Z = 90 and Z = 92 respectively) are actinide elements with similar atomic masses and their abundances are expected to closely correlate. However, few actual U measurements have been made in the Galactic disk. Eu (Z = 63) is a lanthanide element. It is synthesized similarly in the r-process and correlates in lockstep with Th in solar twins after accounting for radioactive decay \citep{Botelho2019}. By the transitive property, we assume that Eu will track with U as well. Herein we use Eu as a direct tracer for U and Th such that the [Eu/U]\footnote{Here we adopt the standard notation where the relative abundance of elements A and B in a given star compared to the sun can be written as $[A/B] = \log_{10}(N_{A}/N_{B})_{\star} - \log_{10}(N_{A}/N_{B})_\sun$.} and [Eu/Th] = 0. This allows us to place constraints on how nearby planets may have evolved by scaling the Eu (and thus U and Th) content of the planet/star compared to the terrestrial value. However, the exact translation between Eu and Th (and U) is not fully described. A recent study by \cite{2022MNRAS.516.3786M} suggests an increase in [Eu/Th] at [Fe/H] $<$ -0.2 and a much larger dispersion in [Eu/Th] at all [Fe/H]. This is in tension with \citet{Botelho2019} who shows no correlation between Eu and Th ([Th/Eu] = 0), and a small scatter at solar metallicity. This is discussed in more detail in Section \ref{sec:Implications on Dynamo Failure Range}. 

Relevant to the above two outstanding questions, 1) how the Galactic disk r-process distribution connects with its synthesis site(s) and 2) how real variations in radioactive element content affect the evolution of nearby planets, we focus on constraining the variation of Eu from star to star in the stellar neighborhood. To address this, we quantify the intrinsic scatter of [Eu/H] at fixed metallicity and temperature, removing systematic trends that artificially inflate the dispersion of Eu. We then turn to existing Galactic chemical models to connect the observable distribution of the r-process with the proposed sites of synthesis. Additionally, equipped with Eu as a proxy for U and Th, we use the dispersion of Eu to predict geodynamo evolution in rocky Earth-similar planets as a function of radiogenic element abundance. A small intrinsic scatter allows us to place meaningful constraints on how these planets might have evolved.

\section{Methodology} \label{sec:methodology}

In this study we adopt chemical abundances from two modern surveys in the literature: 1) the high-resolution (R = 115 000) survey of 1111 F-, G-, and K-type stars observed as a part of the High Accuracy Radial velocity Planet Searcher (HARPS) Guaranteed Time Observations (GTO) program \citep{DelgadoMena2017} (hereafter referred to as DM), and 2) the high-resolution (R = 40 000-110 000) multi-telescope campaign to observe 714 F- and G-type stars in the solar neighborhood, \citet{Battistini2016} and \citet{Bensby2014} (hereafter referred to as BB). Both samples represent extremely high S/N surveys of the local stellar F,G, and K population. The DM analysis measures Eu in 577 stars using the red 6645 \AA absorption line. The BB analysis measures Eu in 377 stars using the blue 4129 \AA absorption line. The DM analysis made chemical measurements by fitting equivalent widths (EW) under a standard LTE analysis with the 2014 version of MOOG \citep{Sneden1973}. Abundance errors were determined empirically for each star as the quadratic sum of EW measurement uncertainty, errors in the atomic parameters of the lines, and uncertainty in the stellar atmospheric parameters. BB analysis made chemical measurements by fitting observed spectra to synthetic spectra calculated by the MARCS2012 code \citep{2008A&A...486..951G} under the assumption of LTE and one-dimensional plane-parallel model atmospheres. The synthesis fitting was completed using {\it Spectroscopy Made Easy} (SME, \citealt{1996A&AS..118..595V,2005ApJS..159..141V}). The final abundances are normalized relative to the average solar value described therein on a line-by-line basis. Random errors are quantified by stochastically varying stellar parameters, and by accounting for line-to-line scatter in abundance measurements for elements other than Eu.
    
Initially, the dispersion of [Eu/H] in both samples appears to be quite large. This is the result of systematic inflation by astrophysical trends related to metallicity/age and Galactic population (measured as [$\alpha$/H]), and methodological biases related to the effective-temperature ($T_{\rm eff}$) dependence of [Eu/H] abundance. To mitigate these biases and recover the true star-to-star dispersion in [Eu/H], we detrend both samples by fitting a linear contour (plane) to each sample: 
\begin{equation} \label{eq:plane}
    {\rm [Eu/H]}_{\rm Detrended}
    = {\rm [Eu/H]} - (A \times T_{eff} +B \times [\alpha/{\rm H}] + C)
\end{equation}
This allows the scatter of [Eu/H] to be evaluated for fixed temperature and metallicity.

\subsection{Systematic Biases in Metallicity - $\alpha$ Elements}\label{sec:Metallicity}

The metallicity dependence on [Eu/H] is a consequence of stellar evolution. Colloquially, metallicity refers to [Fe/H], a well established stellar evolutionary probe. However, in this study, we have chosen to use an $\alpha$-element based metallicity: [$\alpha$/H]. This is motivated by chemical differentiation within the stellar neighborhood in the high-$\alpha$ and low-$\alpha$ groups (correlated with the Galactic thick and thin disks). High-$\alpha$ population stars have systematically larger [Eu/H] abundances than the low-$\alpha$ population stars, making the correlation between [Fe/H] and [Eu/H] distinct for each group. [Fe/H] and [$\alpha$/H] are both well connected to stellar evolution, so both can be used to trace the history of chemical evolution. In this context fitting to [$\alpha$/H] to account for [Eu/H]'s dependence on metallicity serves the distinct advantage of coupling metallicity to the Galactic $\alpha$ population while still being a robust probe for stellar evolution. 

Our term $\alpha$ metallicity is defined as the unweighted average of silicon (Si), magnesium (Mg), titanium (Ti), and calcium (Ca). Si and Mg are typical elements used to describe $\alpha$ element abundance, while Ti and Ca are incorporated to reduce measurement uncertainty by averaging all four elements.

\subsection{Systematic Biases in $T_{\rm eff}$}\label{sec:effective temperature}

The bias in the effective temperature ($T_{\rm eff}$) is not reflective of any astrophysical process; in principle, $T_{\rm eff}$ should not chemically have an effect on [Eu/H] abundance. However, we find the mean [Eu/H] abundance in BB is systematically offset by ~$0.0418\pm0.0103$ dex compared to DM. This bias is likely the result of differing methodological approaches to measuring Eu. By detrending $T_{\rm eff}$ this offset is reduced to $0.0082\pm0.0101$, which is within the individual Eu measurement errors. The $T_{\rm eff}$ term in the final fitting contour is subtle, but nonetheless crucial when comparing abundances between samples discussed in Section \ref{sec:combining data sets}.

\subsection{Detrended Contour Fitting}\label{sec:Surface Polynomial, Residual Fitting}
The identified trends, which inflate the star-to-star dispersion in [Eu/H], are removed by regression of a linear contour fit simultaneously to [$\alpha$/H] and $T_{\rm eff}$ (Equation \ref{eq:plane}). This is done using the IRAF \citep{IRAF1,IRAF2} task \textit{Surfit}, weighted by $1/\sigma_{Error}^2$, under a reduced $\chi$-squared test. The best-fit coefficients are shown in Table \ref{tab:Contour}. 
    
An expression for the detrended [Eu/H] as a function of metallicity and temperature is created by subtracting the contour fit from [Eu/H] for each sample (Equation \ref{eq:plane}), this function represents [Eu/H] at a fixed metallicity and $T_{\rm eff}$. The standard deviation of the detrended abundances of [Eu/H] is 0.071 dex and 0.081 dex in samples by DM and BB, respectively.

\section{Results and Analysis} \label{sec:Results_and_analysis}
\subsection{Combining Data Sets}\label{sec:combining data sets}

Although the two samples from DM and BB represent some of the highest quality chemical measurements available on the stellar neighborhood, there are still inconsistencies in measured [Eu/H] between stars in each sample. In an attempt to reduce these inconsistencies, we create a cross-matched sample of the 68 stars that appear in common between these surveys and average their detrended [Eu/H] expressions. Within the cross-matched sample the standard deviation about the detrended [Eu/H] is reduced to 0.057 dex and 0.065 dex for DM and BB respectively. For the remainder of this paper, we will reference the combined sample and its dispersion because it represents the highest quality sample available with reduced noise compared to the BB and DM samples alone. We include the original data sets in Figure 1 for reference.

\subsection{Intrinsic Dispersion}\label{sec:intrinsic}
Cross matching our samples provides an opportunity to reduce the detrended [Eu/H] uncertainty even further. If the detrended [Eu/H] dispersion is entirely contributed by Poisson noise, we would expect the dispersion about the averaged detrended [Eu/H] to reduce by a factor of $1/\sqrt{2}$. However, we find that this is not the case; the average detrended [Eu/H] scatter reduces to only 0.047 dex when averaged. This would imply that there is a secondary factor that dominates the detrended [Eu/H] spread. We refer to this term as the {\it intrinsic dispersion}, which we determine using the following approach.

To begin with, the total error budget is assumed to be the sum of three terms: uncertainty contributed by Poisson noise, uncertainty contributed by systematic biases, and the independent intrinsic term. Thus, the error budget on the detrended [Eu/H] for both samples can be written as Equation~\ref{eq:1}:
\begin{equation}\label{eq:1}
\sigma_{\rm Total}^2 = \sigma_{\rm Poisson}^2\ + \sigma_{\rm Systematic}^2 + \sigma_{\rm Intrinsic}^2\
\end{equation}
In the creation of the detrended [Eu/H] expression, all significant systematic biases (metallicity and temperature) were removed. Therefore, we conclude that the uncertainty contributed by the symmetric biases is effectively removed. The total error budget is then written as the sum of only Poisson and intrinsic terms. 
When the two samples are averaged, the resulting uncertainty of average detrended [Eu/H] can be written as the average of the two total error budgets of each sample. The intrinsic term should be constant across both samples and thus can be combined.  
\begin{equation}\label{eq:4}
\sigma_{\rm Average}^2 = \frac{\sigma_{\rm Poisson DM}^2 + \sigma_{\rm Poisson BB}^2 + 2\sigma_{\rm Intrinsic}^2}{4}
\end{equation}
Then it is rearranged to solve for the constant intrinsic uncertainty in equation \ref{eq:5}:
\begin{equation}\label{eq:5}
\sigma_{\rm Intrinsic} = \sqrt{2\sigma_{\rm Average}^2 - \frac{\sigma_{\rm PoissonDM}^2 + \sigma_{\rm PoissonBB}^2}{2}}
\end{equation}
Using $\sigma_{\rm PoissonDM} = 0.057$, $\sigma_{\rm PoissonBB}^2 = 0.065$, and $\sigma_{\rm Average} = 0.047$, this results in an intrinsic uncertainty of $\sigma_{\rm Intrinsic} = 0.025$ dex. We believe this to be the intrinsic star-to-star variation in [Eu/H] in the stellar neighborhood, at a fixed metallicity and temperature. The identified dispersion is corroborated by the solar twins analysis by \citet{Bedell2018} and \citet{Spina2018} discussed in the Appendix ~\ref{sec:twins}.  
Outliers in the samples are discussed in the Appendix ~\ref{sec:Outliers}. 

\section{Discussion} \label{sec:Discussion}
\subsection{Implications for r-Process Production}\label{sec:Implications for r-Process Production}\begin{figure*}
    \centering
        \includegraphics[width=\textwidth]{superplot_completeerror.png}
    \caption{ \textbf{(a)} Top panel shows [Eu/$\alpha$] independent of the temperature bias as a function of [$\alpha$/H]. Small \textit{orange} triangles represent data from DM and small \textit{green} triangles represent data from BB. The cross matched sample of 68 stars described in Section \ref{sec:combining data sets} is shown in big \textit{red} squares. Stars from the solar twins analysis by \cite{Spina2018}, \cite{Bedell2018} are shown as big \textit{blue} circles. A \textit{black} star is used to represent the solar abundance value. The \textit{red dashed} line at [Eu/$\alpha$] = 0.06 represents the critical value for an extended dynamo failure given by the model from \cite{Nimmo2020}. Outliers in [Eu/H] abundances discussed in Section \ref{sec:Outliers} are \textit{outlined in black.} \textbf{(b)} The bottom panel shows [Eu/Fe] independent of the temperature bias as a function of [Fe/H]. The legend is the same as in \textbf{(a)}. }
     \label{fig:allplot} 
\end{figure*}

Understanding how Eu is distributed throughout the local stellar population may shed light on the source of r-process element production. The nucleosynthetic pathways and conditions for the production of r-process elements have been understood since \citet{1957RvMP...29..547B} and \citet{1957PASP...69..201C}, but the specific site(s) responsible for the enrichment of the Galactic disk by the r-process is an area of active study \citep{2023A&ARv..31....1A}. The conditions required to synthesize the elements formed via the r-process are theorized to be met in two scenarios. First, a tiny fraction of exotic CC-SNe explosions may meet the conditions to effectively synthesize the r-process due to either a large magnetic energy density around the central compact object \citep[i.e., magnetars:][]{2018ApJ...864..171M} or the presence of a hyperaccreting disk \citep[i.e., collapsars:][]{2019EPJA...55..203S}. Secondly, NS-NS mergers \citep[e.g.,][]{1974ApJ...192L.145L} are a robust nucleosynthetic pathway for the r-process and have been indirectly observed to power the ensuing kilonova \citep[][]{2017Sci...358.1556C} from the gravitational wave discovery GW170817 \citep{2017PhRvL.119p1101A}. To disentangle these distinct pathways, a useful discriminant is the presence of $\alpha$ elements (including Mg, Si, Ca, and Ti), which are produced primarily in CC-SNe \citep{1995ApJS..101..181W,2006ApJ...653.1145K}. By studying the distribution of the r-process elements with respect to the $\alpha$ elements, the production of the r-process elements can be correlated with the CC-SNe rate.

If the galactic injection of the r-process is dominated by exotic CC-SNe, Eu and $\alpha$ will be co-produced. With the relative fraction of exotic CC-SNe to all CC-SNe fixed, this predicts [Eu/$\alpha$] to remain relatively constant with increasing [$\alpha$/H]; in this case, the constant ratio should be representative of the IMF-weighted yield of the ejecta \citep{Macias2018,Kolborg2023}. However, as shown in Figure~\ref{fig:allplot}(a), we see a significant anticorrelation between [Eu/$\alpha$] and [$\alpha$/H] (Pearson Correlation Coefficient = -0.71). Assuming a CC-SNe pathway, the anticorrelation can be explained by a  metallicity-dependent r-process  or a  metallicity-dependent $\alpha$ production in CC-SNe \citep{2006ApJ...653.1145K}. \citet{2006MNRAS.367.1329R} notes a similar anticorrelation between $\alpha$ elements (given as the average of Mg, Si, Ca, and Ti) and Eu for thin disk stars but shows no discernible correlation for thick disk stars. This is in contrast with the two modern surveys used above (DM and BB) and the solar twins analysis \citep{Spina2018} which all show a Eu-$\alpha$ anticorrelation for both thick and thin disk stars.

 If the r-process occurs primarily in NS-NS mergers, the production of Eu is decoupled from the production of $\alpha$ and Fe \citep{Sneden2008,2015ApJ...807..115S}. The anticorrelation between [Eu/$\alpha$] and [$\alpha$/H] suggests that r-process injection from NS-NS mergers decreases with time compared to the production of $\alpha$ elements. This is expected with NS-NS mergers, as they naturally experience a time delay from neutron-star binary formation to merger, which combined with the star formation history yields a significant decrease in the number of NS-NS merger events at high metallicities \citep[e.g., Figure 1 of][]{2015ApJ...807..115S}. The anticorrelation is then representative of the relative rate of exotic CC-SNe compared to NS-NS mergers.

Furthermore, comparing the distribution of Eu to Fe can provide clues to the time delay from formation to merging of NS-NS pairs. A canonical gravitationally driven delay-time distribution of $t^{-1}$ predicts a flat [Eu/Fe] trend with [Fe/H] \citep[e.g.,][]{2019ApJ...875..106C}. However, this is not observed in the data (Figure~\ref{fig:allplot}(b)). This suggests that the delay-time distribution of NS-NS mergers could differ from $t^{-1}$ of type Ia supernovae \citep[e.g.,][]{2019MNRAS.486.2896S} (set by the merger of white dwarf (WD) binaries \citep[e.g.,][]{2012A&A...546A..70T}, which dominates Fe production in the low-$\alpha$ thin disk. If the mass production per event for both NS-NS (for Eu injection) and WD-WD (for Fe injection)  mergers does not vary, then the delay-time distribution of NS-NS mergers needs to occur more promptly in order to explain the [Eu/Fe] decrease with [Fe/H].

Intriguingly, the type Ia supernova luminosity function shows a trend in which older progenitor systems tend to produce slightly brighter supernovae, indicating a potential correlation between the age of the progenitor and its Fe production \citep{2023MNRAS.520.6214W}. If this trend is found to be strong, the anticorrelation between [Eu/$\alpha$] and [$\alpha$/H] can be explained by assuming that the delay-time distributions of the NS-NS mergers and white dwarf binaries are similar but that Fe production increases with time.
 
A caveat is noted regarding including Si, Ti and Ca in our $\alpha$ metallicity term. Si, Ti, and Ca are fractionally produced by type Ia supernovae, while Mg is produced exclusively in CC-SNe \citep{1995ApJS..101..181W}. A more ``pure" $\alpha$ metallicity was tested using just [Mg/H], which we found to produce similar slopes but with a much noisier dispersion. 

\subsection{Implications Regarding Dynamo Failure Duration}\label{sec:Implications on Dynamo Failure Range}

\begin{figure*}
    \centering
    \includegraphics[width=0.9\linewidth]{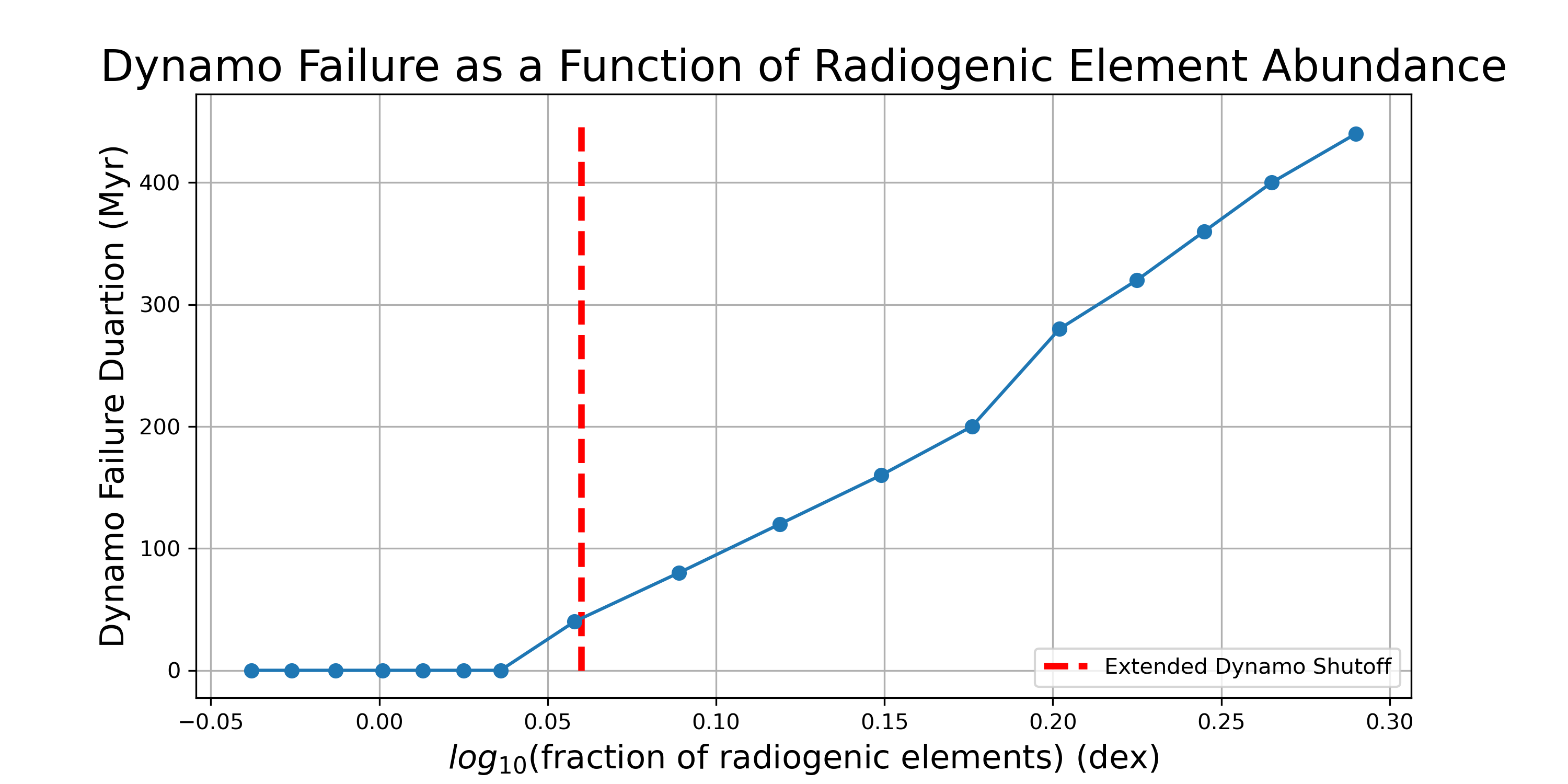}
        \caption{Dynamo failure duration for an Earth-like planet as a function of  terrestrial radiogenic element concentrations (U and Th) relative to $\alpha$ elements. A value of [0] indicates an Earth-like ratio.} See \citet{Nimmo2020} for details on model parameters. A scaled radiogenic fraction of 0.06 dex (red dashed line) results in a dynamo failure of $\sim 40$ Myr, long enough to presumably have devastating effects on surface conditions. This scaled radiogenic fraction threshold corresponds to a stellar abundance of [Eu/$\alpha$] = 0.06 dex.

    \label{fig:dynamo}
\end{figure*}

The decay of radioactive elements in the mantles of rocky planets modulates geodynamic processes, such as a protective planetary magnetic field.  An excessive amount of radioactive heating of the mantle can disrupt the core convection responsible for maintaining a magnetic dynamo \citep{Nimmo2020}. Although the effect of magnetic dynamo failure on atmospheric loss is still debated \citep[e.g.,][]{Gunell2018}, it is generally thought that dynamo persistence is important to prevent the loss of water and ozone and to protect planetary surfaces from bombardment by energetic charged particles \citep[e.g.,][]{Ludin2007}. 

Sustained periods of dynamo failure for just tens of Myr could lead to significant atmospheric loss (particularly if the stellar EUV flux is enhanced) \citep{Lichtenegger2010}. Once the mantle cools, the magnetic dynamos can restart after failure as core convection resumes \citep{Nimmo2020}. However, re-establishment of a planet's atmospheric ${\rm H_{2}O}$ and nitrogen inventory may be more difficult.

 As we have discussed, Th and U are the primary contributors to radiogenic heating at late times. We estimate the abundance of these elements in stars and in the mantles of their rocky planets using the surrogate element Eu compared to the bulk mantle elements Si and Mg. We use an $\alpha$ element metallicity in stars as a proxy for the Si and Mg content of planets because it has the advantage of reducing Poisson noise uncertainties by averaging the abundances [Si/H], [Mg/H], [Ti/H], and [Ca/H]; see Section \ref{sec:Metallicity}.

Relevant to this study are the works by \citet{O'Neill2020} and \citet{Nimmo2020}, in which variations of U and Th concentrations in Earth-like exoplanet mantles and their effect on the generation of a magnetic dynamo are explored. \citet{O'Neill2020} assume variations of U and Th are small between planets and are correlated with the core size. \citet{Nimmo2020} assume larger stochastic variation in Th and U, motivated by the observed scatter in Eu without the detrending corrections discussed above.

Here, we adopt the model described in \citet{Nimmo2020} to predict the content of radioactive elements (quantified by [(U+Th)/Mg]). Using the observed distribution of elements in the disk (traced by [Eu/$\alpha$]), we assess whether the conditions would lead to a persistent dynamo failure. Assuming that the heavy element abundance of stars is directly imprinted on attendant planets, we take Eu as a direct tracer for U and Th ([Th/Eu] = 0; e.g. \citet{Botelho2019}). This model exhibits a dynamo failure of $\sim 40$ Myr at [Eu/$\alpha$] = 0.06 (e.g. Figure \ref{fig:dynamo}). This is a lower limit estimate for a critical period of dynamo collapse in which significant time has passed, allowing substantial atmospheric loss to occur, preventing or delaying the emergence of a hospitable surface environment. For an increasing fraction of radiogenic elements, the period of dynamo collapse increases and presumably so would the consequences. 

Earth-like planets around stars with [Eu/$\alpha$] $\geq 0.06$ are expected to have some degree of extended period of dynamo failure. Figure \ref{fig:allplot}(a) shows that this is true for an increasing fraction of stars with sub-solar $\alpha$-metallicity ([$\alpha$/H]$<-0.25$). In contrast, most stars with [$\alpha$/H]$>-0.25$ do not show an extended period of dynamo failure. At or above solar metallicity ([$\alpha$/H]$\ge0$) all stars in the \citet{Spina2018} and cross-matched samples sit below the critical value of [Eu/$\alpha$] = 0.06, suggesting that such stars lie within a ''habitable metallicity range" where Earth-like planets are unlikely to experience a prolonged duration of dynamo shutdown.  We note that there are stars at or above solar metallicity in the BB and DM samples that sit above the critical value of  [Eu/$\alpha$] = 0.06 but we believe most of these stars are likely 2-3 sigma outliers, i.e. more precise measurements would reveal them to be closer to the mean trend found in the \citet{Spina2018} and combined sample.

 It is important to note that this 1D model has only been constructed for Earth-like planets, with the same radius, mass, and composition as the terrestrial values \citep[e.g.,][]{2007ApJ...669.1279S,2012A&A...547A.112M,2014A&A...561A..41A}. The mass-radius scaling relation for rocky planets has recently been revised in \citet{2020A&A...634A..43O} and \citet{2024A&A...686A.296M}. Furthermore, heat production is expected to decrease at a fixed radius with increasing [Fe/H] due to the larger core-to-mantle ratio \citep{O'Neill2020}. In future work, it would be worth exploring this prospective habitable metallicity range mapped out in radii and composition parameter space based on the changing radiogenic fraction and [Fe/H] ratio \citep[e.g.,][]{2012ApJS..201...15H} using a more comprehensive model that includes a 2D or 3D assessment of core convection. Another complicating factor to consider is tidal heating, which can also affect magnetic field generation \citep{DriscoBarne:2015}. However, this process is exceptionally difficult to constrain given the sensitivity of tidal heating to the interior structure and composition.
 
 We call attention to two caveats regarding our use of Eu as a tracer for Th and U. Firstly, the measurements of Th and Eu in \citet{2022MNRAS.516.3786M} show a much larger dispersion in [Th/Eu] (which may be decreased by the detrending discussed in Section \ref{sec:methodology}) and an increase in [Th/Eu] for [Fe/H] $< -0.2$ dex in contrast to the tight dispersion and [Th/Eu] = 0 at near-solar metallicities, as reported by \citet{Botelho2019}. This increase in Th compared to Eu would exacerbate the degree of radiogenic overheating discussed above, further supporting the notion that near-solar metallicity stars are most likely to support a robust magnetic dynamo on attendant Earth-similar planets. Secondly, Eu is partially produced in the s-process (about $7\%$ at solar metallicity \citep{1999ApJ...525..886A}) unlike U and Th, which are entirely produced in the r-process. The Eu s-process contribution is limited and Eu is still treated as a ``pure r-process" element \citep{2006ApJ...647..685W}. The s-process contribution to Eu in the Sun (about 0.029 dex) is indistinguishable from the abundance errors. It is possible that in the star-to-star variations the fractional contribution of the s-process could be a small contributor to the intrinsic star-to-star scatter we find.

\section{Conclusion}\label{sec:Conclusion}
In conclusion, we have identified metallicity ([$\alpha$/H]) and temperature ($T_{\rm eff}$) as relevant systematic biases inflating the observed scatter of [Eu/H] in the stellar neighborhood. We measured the intrinsic star-to-star scatter of detrended [Eu/H] at a fixed [$\alpha$/H] and $T_{\rm eff}$ to be 0.025 dex, consistent with a chemically well-mixed stellar neighborhood. 

The anticorrelation between [Eu/$\alpha$] and [$\alpha$/H] provides clues to the primary r-process injection site in the Milky Way. Assuming r-process production is dominated by exotic CC-SNe, the observed anticorrelation may be explained by a metallicity-dependent r-process. Alternatively, assuming that the injection of the r-process elements is dominated by NS-NS mergers, the slope of the anticorrelation would give the relative production rate of the r-process elements in NS-NS mergers and $\alpha$ elements in CC-SNe. The observed slope between [Eu/Fe] and [Fe/H] suggests that the delay-time distribution for NS-NS mergers, if responsible for the bulk of the r-process production, is shorter than that of type Ia supernovae. This time delay is inconsistent with the gravitationally driven time delay of $t^{-1}$ set by the merger of WD binaries. Alternatively, the correlation between [Eu/Fe] and [Fe/H] can result from an increase in mass production in WD binaries (supported by brighter type Ia supernovae in elliptical galaxies), even if the time-delay distributions are the same.

Lastly, Th and U are the primary isotopes responsible for radiogenic mantle heating in rocky planets that are several Gyr old. Assuming that planetary compositions reflect their stellar host's atmospheric abundances, we use Eu as a tracer for U and Th and employ a 1D thermal evolutionary model from \citet{Nimmo2020} to constrain dynamo generation. Modeling suggests a significant dynamo failure lasting at least $\geq 40$ Myr beginning at [Eu/$\alpha$] $ = 0.06$ dex. Stars hosting Earth-composition and size planets at or above this critical value are likely to have experienced a dynamo collapse for an extended period of time. This leaves the surface vulnerable to atmospheric stripping and charged-particle bombardment, conditions immediately challenging for a hospitable biosphere. Although the dynamo may resume after sufficient cooling of the mantle, it is unclear if full reestablishment of an atmosphere is possible. 

Below an $\alpha$-metallicity of -0.25 dex, an increasing number of terrestrial planets would suffer dynamo failures during their evolution. This provides a compelling piece of guidance when considering the direction of future observational missions looking for hospitable environments beyond Earth: the r-process abundance and the alpha metallicity of systems should be accounted for as a means to infer radiogenic heating rates and the persistence of a magnetic dynamo. 

\newpage
 
\section*{Acknowledgments} \label{sec:acknowledgments}

This work makes use of the Astrophysics Data System, funded by NASA under Cooperative Agreement 80NSSC21M00561, the SIMBAD database operated at CDS, Strasbourg, France \citep{2000A&AS..143....9W}, the VizieR catalogue access tool, CDS, Strasbourg, France \citep{10.26093/cds/vizier}, and IRAF, distributed by the National Optical Astronomy Observatories \citep{IRAF1,IRAF2}. The original description of the VizieR service was published in \citet{vizier2000}. 

We thank S. M. Faber for fruitful discussions and guidance. ER-R was supported in part by the National Science Foundation (AST-2307710, AST-2206243). EMC was supported in part by the University of California Santa Cruz, Undergraduate Research in Science $\&$ Technology Award. 

\section{Appendix}\label{sec:appendix}
\subsection{Solar Twins Analysis }\label{sec:twins}

Our calculation of an intrinsic dispersion of 0.025 dex in detrended [Eu/H] is supported by the solar twins analysis by \cite{Spina2018} and \citet{Bedell2018}. This sample of 79 stars uses near solar parameters with temperature within $\pm$ 100 K in $T_{\rm eff}$, [Fe/H] within $\pm$ 0.1 dex and surface gravity within $\pm$ 0.1 dex in Log $g$ of the solar value. These restricted parameters combined with high S/N and high resolution contribute to exceptionally low quoted uncertainties on abundance measurements.    
    
We repeat the same simultaneous contour-fit methodology described in Section~\ref{sec:Surface Polynomial, Residual Fitting} to create an expression of detrended [Eu/H] as a function of metallicity and temperature in the solar twins sample. We fit these two terms to [Eu/H] simultaneously in a linear surface fit polynomial with an intercept of -1.534 $\pm$ 0.112 dex, metallicity slope of 0.789 $\pm$ 0.018, and temperature slope of 2.7434$\times 10^{-4}$ $\pm$ 1.944$\times 10^{-5}$ dex, with a reduced $\chi^2$ = 10.22 and standard deviation of 0.031. The reduced $\chi^2$ is notable, a consequence of the exceptionally low uncertainties in elemental abundance measurements.

In our earlier analysis, we were able to cross-match the detrended [Eu/H] measurements in an effort to further reduce the dispersion. However, the sample in \citet{Spina2018} and \citet{Bedell2018} has no stars in common with either DM or BB. Instead, by tweaking the formalism described in Section~\ref{sec:intrinsic} we are able to further reduce the dispersion in the detrended [Eu/H]. Due to the exceptionally low uncertainty in [Eu/H], we claim that the error in [Eu/H] is solely the uncertainty contributed by Poisson noise. In this way, the dispersion of detrended [Eu/H] is given as:
\begin{equation}
    \sigma_{\rm detrended [Eu/H]}^2 = \sigma_{\rm Poisson}^2\ + \sigma_{\rm intrinsic}^2\
    \end{equation}
Rearranging this equation and assuming that the mean error in [Eu/H] is entirely the Poisson noise $\sigma_{\rm Poisson}$ = 0.012 dex, the intrinsic dispersion is determined as $\sigma_{\rm intrinsic}$ = 0.029 dex. 
    
Without a secondary analysis to compare against, we cannot reduce the dispersion any further. The intrinsic dispersion determined here supports the dispersion calculated in Section \ref{sec:intrinsic}. This suggests that we are indeed approaching a star-to-star dispersion reflective of the actual stellar chemical environment intrinsic to the local stellar neighborhood. 

\subsection{Potential outliers}\label{sec:Outliers}

\begin{table*}[ht!]
    \begin{tabular}{|c|c|c|l|} \hline 
         Designation&  Detrended [Eu/H] (dex)& Standard Deviations &Sample\\ \hline \hline
         HD 85512 &  0.244&  3.459&Delgado Mena\\ \hline 
         HD 218511&  0.259&  3.659&Delgado Mena\\ \hline 
         BD+062932&  0.253&  3.579&Delgado Mena\\ \hline 
         HD 73267&  -0.223&  -3.153&Delgado Mena\\ \hline 
         HD 190954&  -0.249&  -3.522&Delgado Mena\\ \hline 
         HD 167677&  0.258&  3.654&Delgado Mena\\ \hline 
         HD 209458&  0.218&  3.079&Delgado Mena\\ \hline 
         HD 148156&  0.255&  3.612&Delgado Mena\\ \hline 
         HD 205591& 0.229& 3.237&Delgado Mena\\ \hline 
         HIP 73385& 0.273& 3.375&Battistini \& Bensby\\ \hline
         HIP 81041& -0.279& -3.447&Battistini \& Bensby\\ \hline
         HIP 81461& 0.270& 3.329&Battistini \& Bensby\\ \hline
         HIP 95106& 0.301& 3.722&Battistini \& Bensby\\ \hline
 \textbf{HD 11397}& \textbf{0.156}&  \textbf{3.343} & \textbf{Cross-matched Sample} \\
 \hline
    \end{tabular}
    \caption{Shown are the 9 stars from DM and 4 stars from BB with detrended [Eu/H] abundance $>$3 SD. Stars only appearing in a single sample use detrended [Eu/H] formulae and SD described in Section \ref{sec:Surface Polynomial, Residual Fitting}. Stars in the cross-matched sample use an averaged detrended [Eu/H] formulae and SD described in Section~\ref{sec:combining data sets}. The last bold entry represents the single star from the cross-matched sample to be $>$ 3 SD. No stars were determined as outliers from the solar twins analysis by \citet{Spina2018} and \citet{Bedell2018}.}
    \label{tab:outliers}
\end{table*}

Within the samples used, a handful of outliers have been identified in [Eu/H] abundance. We consider an outlier as any star whose detrended [Eu/H] value is greater than 3 standard deviations (SD) from the mean. These stars are worth particular attention for two reasons, their unusual [Eu/H] abundance contributes to the inflation of the intrinsic dispersion significantly and secondly these stars may have unique chemical-evolutionary histories worth investigating. 
    
In the sample from DM, which contains 577 stars with [Eu/H] abundance measurements, the standard deviation was found to be 0.071 (methodology described in \ref{sec:Surface Polynomial, Residual Fitting}). Under this standard deviation, 9 stars are measured to have detrended [Eu/H] abundances greater than 3 SD from the mean. These stars are listed in Table \ref{tab:outliers}. Assuming the sample follows a Gaussian distribution, we would only expect 1-2 stars above 3 SD in a sample size of 577 stars; this suggests the existence of at most 7 statistical outliers in detrended [Eu/H]. 

Using a similar analysis, in the sample from BB, containing 377 stars with valid [Eu/H] and [$\alpha$/H] abundance measurements, the standard deviation is found to be 0.081. Using this standard deviation, 4 stars are measured with detrended [Eu/H] of greater than 3 SD. These stars are listed in Table~\ref{tab:outliers}. Following a Gaussian distribution, we only expect one star above 3 SD in a sample size of 377, this suggests the existence of at most two statistical outliers in detrended [Eu/H]. 

No outliers were found in the sample from \citet{Spina2018} and \citet{Bedell2018}. This may be due to the small sample size and/or the exceptionally small quoted uncertainties.

Among the cross-matched sample, there is one outlier, HD 11397, with a detrended [Eu/H] abundance of 0.150 dex: a 3.192 SDs outlier from the mean trend (see Table~\ref{tab:outliers} in bold). HD 11397 is likely a real outlier because its unusual detrended [Eu/H] value is corroborated by both data sets. The cross-matched sample has a significantly lower SD than in the full samples, due to reduced uncertainties from averaging. Consequently, HD 11397, which did not appear as an outlier in the individual samples, now appears as such in the cross-matched sample. Unfortunately, none of the outliers in the individual sample exists in the cross-matched sample.
  
While HD 11397 has corroborating outlier measurements in the cross-matched sample, the anonymously high number of outliers in individual samples drop out of the cross-matched sample and are inconclusive. It is unclear whether these anomalous Eu abundances are a result of measurement error or whether they are representative of r-process enhancement. A homogeneous reanalysis of outlier stars using both absorption lines is recommended to determine the validity of the outlier abundances. HD 11397 is the only star with corroborating Eu measurements, which suggests this is likely an r-process enhanced star. The stars' odd compositions may be a result of a particularly r-process rich birth cloud or another heavy element doping mechanism such as a binary-supernova pair. More analysis is needed to determine the unique properties and histories of the stars listed.

\bibliography{Referances}{}

\begin{thebibliography}{}
\expandafter\ifx\csname natexlab\endcsname\relax\def\natexlab#1{#1}\fi
\providecommand{\url}[1]{\href{#1}{#1}}
\providecommand{\dodoi}[1]{doi:~\href{http://doi.org/#1}{\nolinkurl{#1}}}
\providecommand{\doeprint}[1]{\href{http://ascl.net/#1}{\nolinkurl{http://ascl.net/#1}}}
\providecommand{\doarXiv}[1]{\href{https://arxiv.org/abs/#1}{\nolinkurl{https://arxiv.org/abs/#1}}}

\bibitem[{{Abbott} {et~al.}(2017){Abbott}, {Abbott}, {Abbott}, {Acernese}, {Ackley}, {Adams}, {Adams}, {Addesso}, {Adhikari}, {Adya}, {Affeldt}, {Afrough}, {Agarwal}, {Agathos}, {Agatsuma}, {Aggarwal}, {Aguiar}, {Aiello}, {Ain}, {Ajith}, {Allen}, {Allen}, {Allocca}, {Altin}, {Amato}, {Ananyeva}, {Anderson}, {Anderson}, {Angelova}, {Antier}, {Appert}, {Arai}, {Araya}, {Areeda}, {Arnaud}, {Arun}, {Ascenzi}, {Ashton}, {Ast}, {Aston}, {Astone}, {Atallah}, {Aufmuth}, {Aulbert}, {AultONeal}, {Austin}, {Avila-Alvarez}, {Babak}, {Bacon}, {Bader}, {Bae}, {Bailes}, {Baker}, {Baldaccini}, {Ballardin}, {Ballmer}, {Banagiri}, {Barayoga}, {Barclay}, {Barish}, {Barker}, {Barkett}, {Barone}, {Barr}, {Barsotti}, {Barsuglia}, {Barta}, {Barthelmy}, {Bartlett}, {Bartos}, {Bassiri}, {Basti}, {Batch}, {Bawaj}, {Bayley}, {Bazzan}, {B{\'e}csy}, {Beer}, {Bejger}, {Belahcene}, {Bell}, {Berger}, {Bergmann}, {Bernuzzi}, {Bero}, {Berry}, {Bersanetti}, {Bertolini}, {Betzwieser}, {Bhagwat}, {Bhandare}, {Bilenko}, {Billingsley}, {Billman},
  {Birch}, {Birney}, {Birnholtz}, {Biscans}, {Biscoveanu}, {Bisht}, {Bitossi}, {Biwer}, {Bizouard}, {Blackburn}, {Blackman}, {Blair}, {Blair}, {Blair}, {Bloemen}, {Bock}, {Bode}, {Boer}, {Bogaert}, {Bohe}, {Bondu}, {Bonilla}, {Bonnand}, {Boom}, {Bork}, {Boschi}, {Bose}, {Bossie}, {Bouffanais}, {Bozzi}, {Bradaschia}, {Brady}, {Branchesi}, {Brau}, {Briant}, {Brillet}, {Brinkmann}, {Brisson}, {Brockill}, {Broida}, {Brooks}, {Brown}, {Brown}, {Brunett}, {Buchanan}, {Buikema}, {Bulik}, {Bulten}, {Buonanno}, {Buskulic}, {Buy}, {Byer}, {Cabero}, {Cadonati}, {Cagnoli}, {Cahillane}, {Calder{\'o}n Bustillo}, {Callister}, {Calloni}, {Camp}, {Canepa}, {Canizares}, {Cannon}, {Cao}, {Cao}, {Capano}, {Capocasa}, {Carbognani}, {Caride}, {Carney}, {Carullo}, {Casanueva Diaz}, {Casentini}, {Caudill}, {Cavagli{\`a}}, {Cavalier}, {Cavalieri}, {Cella}, {Cepeda}, {Cerd{\'a}-Dur{\'a}n}, {Cerretani}, {Cesarini}, {Chamberlin}, {Chan}, {Chao}, {Charlton}, {Chase}, {Chassande-Mottin}, {Chatterjee}, {Chatziioannou}, {Cheeseboro},
  {Chen}, {Chen}, {Chen}, {Cheng}, {Chia}, {Chincarini}, {Chiummo}, {Chmiel}, {Cho}, {Cho}, {Chow}, {Christensen}, {Chu}, {Chua}, {Chua}, {Chung}, {Chung}, {Ciani}, {Ciolfi}, {Cirelli}, {Cirone}, {Clara}, {Clark}, {Clearwater}, {Cleva}, {Cocchieri}, {Coccia}, {Cohadon}, {Cohen}, {Colla}, {Collette}, {Cominsky}, {Constancio}, {Conti}, {Cooper}, {Corban}, {Corbitt}, {Cordero-Carri{\'o}n}, {Corley}, {Cornish}, {Corsi}, {Cortese}, {Costa}, {Coughlin}, {Coughlin}, {Coulon}, {Countryman}, {Couvares}, {Covas}, {Cowan}, {Coward}, {Cowart}, {Coyne}, {Coyne}, {Creighton}, {Creighton}, {Cripe}, {Crowder}, {Cullen}, {Cumming}, {Cunningham}, {Cuoco}, {Dal Canton}, {D{\'a}lya}, {Danilishin}, {D'Antonio}, {Danzmann}, {Dasgupta}, {Da Silva Costa}, {Dattilo}, {Dave}, {Davier}, {Davis}, {Daw}, {Day}, {De}, {DeBra}, {Degallaix}, {De Laurentis}, {Del{\'e}glise}, {Del Pozzo}, {Demos}, {Denker}, {Dent}, {De Pietri}, {Dergachev}, {De Rosa}, {DeRosa}, {De Rossi}, {DeSalvo}, {de Varona}, {Devenson}, {Dhurandhar}, {D{\'\i}az},
  {Dietrich}, {Di Fiore}, {Di Giovanni}, {Di Girolamo}, {Di Lieto}, {Di Pace}, {Di Palma}, {Di Renzo}, {Doctor}, {Dolique}, {Donovan}, {Dooley}, {Doravari}, {Dorrington}, {Douglas}, {Dovale {\'A}lvarez}, {Downes}, {Drago}, {Dreissigacker}, {Driggers}, {Du}, {Ducrot}, {Dudi}, {Dupej}, {Dwyer}, {Edo}, {Edwards}, {Effler}, {Eggenstein}, {Ehrens}, {Eichholz}, {Eikenberry}, {Eisenstein}, {Essick}, {Estevez}, {Etienne}, {Etzel}, {Evans}, {Evans}, {Factourovich}, {Fafone}, {Fair}, {Fairhurst}, {Fan}, {Farinon}, {Farr}, {Farr}, {Fauchon-Jones}, {Favata}, {Fays}, {Fee}, {Fehrmann}, {Feicht}, {Fejer}, {Fernandez-Galiana}, {Ferrante}, {Ferreira}, {Ferrini}, {Fidecaro}, {Finstad}, {Fiori}, {Fiorucci}, {Fishbach}, {Fisher}, {Fitz-Axen}, {Flaminio}, {Fletcher}, {Fong}, {Font}, {Forsyth}, {Forsyth}, {Fournier}, {Frasca}, {Frasconi}, {Frei}, {Freise}, {Frey}, {Frey}, {Fries}, {Fritschel}, {Frolov}, {Fulda}, {Fyffe}, {Gabbard}, {Gadre}, {Gaebel}, {Gair}, {Gammaitoni}, {Ganija}, {Gaonkar}, {Garcia-Quiros}, {Garufi}, {Gateley},
  {Gaudio}, {Gaur}, {Gayathri}, {Gehrels}, {Gemme}, {Genin}, {Gennai}, {George}, {George}, {Gergely}, {Germain}, {Ghonge}, {Ghosh}, {Ghosh}, {Ghosh}, {Giaime}, {Giardina}, {Giazotto}, {Gill}, {Glover}, {Goetz}, {Goetz}, {Gomes}, {Goncharov}, {Gonz{\'a}lez}, {Gonzalez Castro}, {Gopakumar}, {Gorodetsky}, {Gossan}, {Gosselin}, {Gouaty}, {Grado}, {Graef}, {Granata}, {Grant}, {Gras}, {Gray}, {Greco}, {Green}, {Gretarsson}, {Groot}, {Grote}, {Grunewald}, {Gruning}, {Guidi}, {Guo}, {Gupta}, {Gupta}, {Gushwa}, {Gustafson}, {Gustafson}, {Halim}, {Hall}, {Hall}, {Hamilton}, {Hammond}, {Haney}, {Hanke}, {Hanks}, {Hanna}, {Hannam}, {Hannuksela}, {Hanson}, {Hardwick}, {Harms}, {Harry}, {Harry}, {Hart}, {Haster}, {Haughian}, {Healy}, {Heidmann}, {Heintze}, {Heitmann}, {Hello}, {Hemming}, {Hendry}, {Heng}, {Hennig}, {Heptonstall}, {Heurs}, {Hild}, {Hinderer}, {Ho}, {Hoak}, {Hofman}, {Holt}, {Holz}, {Hopkins}, {Horst}, {Hough}, {Houston}, {Howell}, {Hreibi}, {Hu}, {Huerta}, {Huet}, {Hughey}, {Husa}, {Huttner}, {Huynh-Dinh},
  {Indik}, {Inta}, {Intini}, {Isa}, {Isac}, {Isi}, {Iyer}, {Izumi}, {Jacqmin}, {Jani}, {Jaranowski}, {Jawahar}, {Jim{\'e}nez-Forteza}, {Johnson}, {Johnson-McDaniel}, {Jones}, {Jones}, {Jonker}, {Ju}, {Junker}, {Kalaghatgi}, {Kalogera}, {Kamai}, {Kandhasamy}, {Kang}, {Kanner}, {Kapadia}, {Karki}, {Karvinen}, {Kasprzack}, {Kastaun}, {Katolik}, {Katsavounidis}, {Katzman}, {Kaufer}, {Kawabe}, {K{\'e}f{\'e}lian}, {Keitel}, {Kemball}, {Kennedy}, {Kent}, {Key}, {Khalili}, {Khan}, {Khan}, {Khan}, {Khazanov}, {Kijbunchoo}, {Kim}, {Kim}, {Kim}, {Kim}, {Kim}, {Kim}, {Kimbrell}, {King}, {King}, {Kinley-Hanlon}, {Kirchhoff}, {Kissel}, {Kleybolte}, {Klimenko}, {Knowles}, {Koch}, {Koehlenbeck}, {Koley}, {Kondrashov}, {Kontos}, {Korobko}, {Korth}, {Kowalska}, {Kozak}, {Kr{\"a}mer}, {Kringel}, {Krishnan}, {Kr{\'o}lak}, {Kuehn}, {Kumar}, {Kumar}, {Kumar}, {Kuo}, {Kutynia}, {Kwang}, {Lackey}, {Lai}, {Landry}, {Lang}, {Lange}, {Lantz}, {Lanza}, {Larson}, {Lartaux-Vollard}, {Lasky}, {Laxen}, {Lazzarini}, {Lazzaro}, {Leaci},
  {Leavey}, {Lee}, {Lee}, {Lee}, {Lee}, {Lee}, {Lehmann}, {Lenon}, {Leon}, {Leonardi}, {Leroy}, {Letendre}, {Levin}, {Li}, {Linker}, {Littenberg}, {Liu}, {Liu}, {Lo}, {Lockerbie}, {London}, {Lord}, {Lorenzini}, {Loriette}, {Lormand}, {Losurdo}, {Lough}, {Lousto}, {Lovelace}, {L{\"u}ck}, {Lumaca}, {Lundgren}, {Lynch}, {Ma}, {Macas}, {Macfoy}, {Machenschalk}, {MacInnis}, {Macleod}, {Maga{\~n}a Hernandez}, {Maga{\~n}a-Sandoval}, {Maga{\~n}a Zertuche}, {Magee}, {Majorana}, {Maksimovic}, {Man}, {Mandic}, {Mangano}, {Mansell}, {Manske}, {Mantovani}, {Marchesoni}, {Marion}, {M{\'a}rka}, {M{\'a}rka}, {Markakis}, {Markosyan}, {Markowitz}, {Maros}, {Marquina}, {Marsh}, {Martelli}, {Martellini}, {Martin}, {Martin}, {Martynov}, {Marx}, {Mason}, {Massera}, {Masserot}, {Massinger}, {Masso-Reid}, {Mastrogiovanni}, {Matas}, {Matichard}, {Matone}, {Mavalvala}, {Mazumder}, {McCarthy}, {McClelland}, {McCormick}, {McCuller}, {McGuire}, {McIntyre}, {McIver}, {McManus}, {McNeill}, {McRae}, {McWilliams}, {Meacher}, {Meadors},
  {Mehmet}, {Meidam}, {Mejuto-Villa}, {Melatos}, {Mendell}, {Mercer}, {Merilh}, {Merzougui}, {Meshkov}, {Messenger}, {Messick}, {Metzdorff}, {Meyers}, {Miao}, {Michel}, {Middleton}, {Mikhailov}, {Milano}, {Miller}, {Miller}, {Miller}, {Millhouse}, {Milovich-Goff}, {Minazzoli}, {Minenkov}, {Ming}, {Mishra}, {Mitra}, {Mitrofanov}, {Mitselmakher}, {Mittleman}, {Moffa}, {Moggi}, {Mogushi}, {Mohan}, {Mohapatra}, {Molina}, {Montani}, {Moore}, {Moraru}, {Moreno}, {Morisaki}, {Morriss}, {Mours}, {Mow-Lowry}, {Mueller}, {Muir}, {Mukherjee}, {Mukherjee}, {Mukherjee}, {Mukund}, {Mullavey}, {Munch}, {Mu{\~n}iz}, {Muratore}, {Murray}, {Nagar}, {Napier}, {Nardecchia}, {Naticchioni}, {Nayak}, {Neilson}, {Nelemans}, {Nelson}, {Nery}, {Neunzert}, {Nevin}, {Newport}, {Newton}, {Ng}, {Nguyen}, {Nguyen}, {Nichols}, {Nielsen}, {Nissanke}, {Nitz}, {Noack}, {Nocera}, {Nolting}, {North}, {Nuttall}, {Oberling}, {O'Dea}, {Ogin}, {Oh}, {Oh}, {Ohme}, {Okada}, {Oliver}, {Oppermann}, {Oram}, {O'Reilly}, {Ormiston}, {Ortega},
  {O'Shaughnessy}, {Ossokine}, {Ottaway}, {Overmier}, {Owen}, {Pace}, {Page}, {Page}, {Pai}, {Pai}, {Palamos}, {Palashov}, {Palomba}, {Pal-Singh}, {Pan}, {Pan}, {Pang}, {Pang}, {Pankow}, {Pannarale}, {Pant}, {Paoletti}, {Paoli}, {Papa}, {Parida}, {Parker}, {Pascucci}, {Pasqualetti}, {Passaquieti}, {Passuello}, {Patil}, {Patricelli}, {Pearlstone}, {Pedraza}, {Pedurand}, {Pekowsky}, {Pele}, {Penn}, {Perez}, {Perreca}, {Perri}, {Pfeiffer}, {Phelps}, {Piccinni}, {Pichot}, {Piergiovanni}, {Pierro}, {Pillant}, {Pinard}, {Pinto}, {Pirello}, {Pitkin}, {Poe}, {Poggiani}, {Popolizio}, {Porter}, {Post}, {Powell}, {Prasad}, {Pratt}, {Pratten}, {Predoi}, {Prestegard}, {Prijatelj}, {Principe}, {Privitera}, {Prix}, {Prodi}, {Prokhorov}, {Puncken}, {Punturo}, {Puppo}, {P{\"u}rrer}, {Qi}, {Quetschke}, {Quintero}, {Quitzow-James}, {Raab}, {Rabeling}, {Radkins}, {Raffai}, {Raja}, {Rajan}, {Rajbhandari}, {Rakhmanov}, {Ramirez}, {Ramos-Buades}, {Rapagnani}, {Raymond}, {Razzano}, {Read}, {Regimbau}, {Rei}, {Reid}, {Reitze}, {Ren},
  {Reyes}, {Ricci}, {Ricker}, {Rieger}, {Riles}, {Rizzo}, {Robertson}, {Robie}, {Robinet}, {Rocchi}, {Rolland}, {Rollins}, {Roma}, {Romano}, {Romano}, {Romel}, {Romie}, {Rosi{\'n}ska}, {Ross}, {Rowan}, {R{\"u}diger}, {Ruggi}, {Rutins}, {Ryan}, {Sachdev}, {Sadecki}, {Sadeghian}, {Sakellariadou}, {Salconi}, {Saleem}, {Salemi}, {Samajdar}, {Sammut}, {Sampson}, {Sanchez}, {Sanchez}, {Sanchis-Gual}, {Sandberg}, {Sanders}, {Sassolas}, {Sathyaprakash}, {Saulson}, {Sauter}, {Savage}, {Sawadsky}, {Schale}, {Scheel}, {Scheuer}, {Schmidt}, {Schmidt}, {Schnabel}, {Schofield}, {Sch{\"o}nbeck}, {Schreiber}, {Schuette}, {Schulte}, {Schutz}, {Schwalbe}, {Scott}, {Scott}, {Seidel}, {Sellers}, {Sengupta}, {Sentenac}, {Sequino}, {Sergeev}, {Shaddock}, {Shaffer}, {Shah}, {Shahriar}, {Shaner}, {Shao}, {Shapiro}, {Shawhan}, {Sheperd}, {Shoemaker}, {Shoemaker}, {Siellez}, {Siemens}, {Sieniawska}, {Sigg}, {Silva}, {Singer}, {Singh}, {Singhal}, {Sintes}, {Slagmolen}, {Smith}, {Smith}, {Smith}, {Somala}, {Son}, {Sonnenberg}, {Sorazu},
  {Sorrentino}, {Souradeep}, {Spencer}, {Srivastava}, {Staats}, {Staley}, {Steinke}, {Steinlechner}, {Steinlechner}, {Steinmeyer}, {Stevenson}, {Stone}, {Stops}, {Strain}, {Stratta}, {Strigin}, {Strunk}, {Sturani}, {Stuver}, {Summerscales}, {Sun}, {Sunil}, {Suresh}, {Sutton}, {Swinkels}, {Szczepa{\'n}czyk}, {Tacca}, {Tait}, {Talbot}, {Talukder}, {Tanner}, {T{\'a}pai}, {Taracchini}, {Tasson}, {Taylor}, {Taylor}, {Tewari}, {Theeg}, {Thies}, {Thomas}, {Thomas}, {Thomas}, {Thorne}, {Thorne}, {Thrane}, {Tiwari}, {Tiwari}, {Tokmakov}, {Toland}, {Tonelli}, {Tornasi}, {Torres-Forn{\'e}}, {Torrie}, {T{\"o}yr{\"a}}, {Travasso}, {Traylor}, {Trinastic}, {Tringali}, {Trozzo}, {Tsang}, {Tse}, {Tso}, {Tsukada}, {Tsuna}, {Tuyenbayev}, {Ueno}, {Ugolini}, {Unnikrishnan}, {Urban}, {Usman}, {Vahlbruch}, {Vajente}, {Valdes}, {Vallisneri}, {van Bakel}, {van Beuzekom}, {van den Brand}, {Van Den Broeck}, {Vander-Hyde}, {van der Schaaf}, {van Heijningen}, {van Veggel}, {Vardaro}, {Varma}, {Vass}, {Vas{\'u}th}, {Vecchio}, {Vedovato},
  {Veitch}, {Veitch}, {Venkateswara}, {Venugopalan}, {Verkindt}, {Vetrano}, {Vicer{\'e}}, {Viets}, {Vinciguerra}, {Vine}, {Vinet}, {Vitale}, {Vo}, {Vocca}, {Vorvick}, {Vyatchanin}, {Wade}, {Wade}, {Wade}, {Walet}, {Walker}, {Wallace}, {Walsh}, {Wang}, {Wang}, {Wang}, {Wang}, {Wang}, {Ward}, {Warner}, {Was}, {Watchi}, {Weaver}, {Wei}, {Weinert}, {Weinstein}, {Weiss}, {Wen}, {Wessel}, {We{\ss}els}, {Westerweck}, {Westphal}, {Wette}, {Whelan}, {Whitcomb}, {Whiting}, {Whittle}, {Wilken}, {Williams}, {Williams}, {Williamson}, {Willis}, {Willke}, {Wimmer}, {Winkler}, {Wipf}, {Wittel}, {Woan}, {Woehler}, {Wofford}, {Wong}, {Worden}, {Wright}, {Wu}, {Wysocki}, {Xiao}, {Yamamoto}, {Yancey}, {Yang}, {Yap}, {Yazback}, {Yu}, {Yu}, {Yvert}, {Zadro{\.Z}ny}, {Zanolin}, {Zelenova}, {Zendri}, {Zevin}, {Zhang}, {Zhang}, {Zhang}, {Zhang}, {Zhao}, {Zhou}, {Zhou}, {Zhu}, {Zhu}, {Zimmerman}, {Zucker}, {Zweizig}, {LIGO Scientific Collaboration}, \& {Virgo Collaboration}}]{2017PhRvL.119p1101A}
{Abbott}, B.~P., {Abbott}, R., {Abbott}, T.~D., {et~al.} 2017, \prl, 119, 161101, \dodoi{10.1103/PhysRevLett.119.161101}

\bibitem[{{Alibert}(2014)}]{2014A&A...561A..41A}
{Alibert}, Y. 2014, \aap, 561, A41, \dodoi{10.1051/0004-6361/201322293}

\bibitem[{{Arcones} \& {Thielemann}(2023)}]{2023A&ARv..31....1A}
{Arcones}, A., \& {Thielemann}, F.-K. 2023, \aapr, 31, 1, \dodoi{10.1007/s00159-022-00146-x}

\bibitem[{{Arlandini} {et~al.}(1999){Arlandini}, {K{\"a}ppeler}, {Wisshak}, {Gallino}, {Lugaro}, {Busso}, \& {Straniero}}]{1999ApJ...525..886A}
{Arlandini}, C., {K{\"a}ppeler}, F., {Wisshak}, K., {et~al.} 1999, \apj, 525, 886, \dodoi{10.1086/307938}

\bibitem[{{Battistini} \& {Bensby}(2016)}]{Battistini2016}
{Battistini}, C., \& {Bensby}, T. 2016, \aap, 586, A49, \dodoi{10.1051/0004-6361/201527385}

\bibitem[{{Bedell} {et~al.}(2018){Bedell}, {Bean}, {Mel{\'e}ndez}, {Spina}, {Ram{\'\i}rez}, {Asplund}, {Alves-Brito}, {dos Santos}, {Dreizler}, {Yong}, {Monroe}, \& {Casagrande}}]{Bedell2018}
{Bedell}, M., {Bean}, J.~L., {Mel{\'e}ndez}, J., {et~al.} 2018, \apj, 865, 68, \dodoi{10.3847/1538-4357/aad908}

\bibitem[{{Bensby} {et~al.}(2005){Bensby}, {Feltzing}, {Lundstr{\"o}m}, \& {Ilyin}}]{2005A&A...433..185B}
{Bensby}, T., {Feltzing}, S., {Lundstr{\"o}m}, I., \& {Ilyin}, I. 2005, \aap, 433, 185, \dodoi{10.1051/0004-6361:20040332}

\bibitem[{{Bensby} {et~al.}(2014){Bensby}, {Feltzing}, \& {Oey}}]{Bensby2014}
{Bensby}, T., {Feltzing}, S., \& {Oey}, M.~S. 2014, \aap, 562, A71, \dodoi{10.1051/0004-6361/201322631}

\bibitem[{Bono {et~al.}(2019)Bono, Tarduno, Nimmo, \& Cottrell}]{Bono-etal:2019}
Bono, R.~K., Tarduno, J.~A., Nimmo, F., \& Cottrell, R.~D. 2019, Nature Geoscience, 12, 143

\bibitem[{{Botelho} {et~al.}(2019){Botelho}, {Milone}, {Mel{\'e}ndez}, {Bedell}, {Spina}, {Asplund}, {dos Santos}, {Bean}, {Ram{\'\i}rez}, {Yong}, {Dreizler}, {Alves-Brito}, \& {Yana Galarza}}]{Botelho2019}
{Botelho}, R.~B., {Milone}, A. d.~C., {Mel{\'e}ndez}, J., {et~al.} 2019, \mnras, 482, 1690, \dodoi{10.1093/mnras/sty2791}

\bibitem[{{Boujibar} {et~al.}(2020){Boujibar}, {Driscoll}, \& {Fei}}]{Boujibar2020}
{Boujibar}, A., {Driscoll}, P., \& {Fei}, Y. 2020, Journal of Geophysical Research (Planets), 125, e06124, \dodoi{10.1029/2019JE006124}

\bibitem[{{Burbidge} {et~al.}(1957){Burbidge}, {Burbidge}, {Fowler}, \& {Hoyle}}]{1957RvMP...29..547B}
{Burbidge}, E.~M., {Burbidge}, G.~R., {Fowler}, W.~A., \& {Hoyle}, F. 1957, Reviews of Modern Physics, 29, 547, \dodoi{10.1103/RevModPhys.29.547}

\bibitem[{{Cameron}(1957)}]{1957PASP...69..201C}
{Cameron}, A.~G.~W. 1957, \pasp, 69, 201, \dodoi{10.1086/127051}

\bibitem[{{Cescutti} {et~al.}(2005){Cescutti}, {Fran{\c{c}}ois}, \& {Matteucci}}]{Cescutti2005}
{Cescutti}, G., {Fran{\c{c}}ois}, P., \& {Matteucci}, F. 2005, in IAU Symposium, Vol. 228, From Lithium to Uranium: Elemental Tracers of Early Cosmic Evolution, ed. V.~{Hill}, P.~{Francois}, \& F.~{Primas}, 445--450, \dodoi{10.1017/S1743921305006198}

\bibitem[{{Chen} {et~al.}(2024){Chen}, {Li}, {Chen}, {Hu}, \& {Liang}}]{Chen2024}
{Chen}, M.-H., {Li}, L.-X., {Chen}, Q.-H., {Hu}, R.-C., \& {Liang}, E.-W. 2024, \mnras, 529, 1154, \dodoi{10.1093/mnras/stae475}

\bibitem[{{C{\^o}t{\'e}} {et~al.}(2019){C{\^o}t{\'e}}, {Eichler}, {Arcones}, {Hansen}, {Simonetti}, {Frebel}, {Fryer}, {Pignatari}, {Reichert}, {Belczynski}, \& {Matteucci}}]{2019ApJ...875..106C}
{C{\^o}t{\'e}}, B., {Eichler}, M., {Arcones}, A., {et~al.} 2019, \apj, 875, 106, \dodoi{10.3847/1538-4357/ab10db}

\bibitem[{{Coulter} {et~al.}(2017){Coulter}, {Foley}, {Kilpatrick}, {Drout}, {Piro}, {Shappee}, {Siebert}, {Simon}, {Ulloa}, {Kasen}, {Madore}, {Murguia-Berthier}, {Pan}, {Prochaska}, {Ramirez-Ruiz}, {Rest}, \& {Rojas-Bravo}}]{2017Sci...358.1556C}
{Coulter}, D.~A., {Foley}, R.~J., {Kilpatrick}, C.~D., {et~al.} 2017, Science, 358, 1556, \dodoi{10.1126/science.aap9811}

\bibitem[{{Dean} \& {Fern{\'a}ndez}(2024)}]{2024PhRvD.110h3024D}
{Dean}, C., \& {Fern{\'a}ndez}, R. 2024, \prd, 110, 083024, \dodoi{10.1103/PhysRevD.110.083024}

\bibitem[{{Delgado Mena} {et~al.}(2017){Delgado Mena}, {Tsantaki}, {Adibekyan}, {Sousa}, {Santos}, {Gonz{\'a}lez Hern{\'a}ndez}, \& {Israelian}}]{DelgadoMena2017}
{Delgado Mena}, E., {Tsantaki}, M., {Adibekyan}, V.~Z., {et~al.} 2017, \aap, 606, A94, \dodoi{10.1051/0004-6361/201730535}

\bibitem[{Driscoll \& Barnes(2015)}]{DriscoBarne:2015}
Driscoll, P.~E., \& Barnes, R. 2015, Astrobiology, 15, 739

\bibitem[{{Foley} \& {Smye}(2018)}]{Foley2018}
{Foley}, B.~J., \& {Smye}, A.~J. 2018, Astrobiology, 18, 873, \dodoi{10.1089/ast.2017.1695}

\bibitem[{{Frank} {et~al.}(2014){Frank}, {Meyer}, \& {Mojzsis}}]{Frank2014}
{Frank}, E.~A., {Meyer}, B.~S., \& {Mojzsis}, S.~J. 2014, \icarus, 243, 274, \dodoi{10.1016/j.icarus.2014.08.031}

\bibitem[{{Guiglion} {et~al.}(2018){Guiglion}, {de Laverny}, {Recio-Blanco}, \& {Prantzos}}]{2018A&A...619A.143G}
{Guiglion}, G., {de Laverny}, P., {Recio-Blanco}, A., \& {Prantzos}, N. 2018, \aap, 619, A143, \dodoi{10.1051/0004-6361/201833782}

\bibitem[{{Gunell} {et~al.}(2018){Gunell}, {Maggiolo}, {Nilsson}, {Stenberg Wieser}, {Slapak}, {Lindkvist}, {Hamrin}, \& {De Keyser}}]{Gunell2018}
{Gunell}, H., {Maggiolo}, R., {Nilsson}, H., {et~al.} 2018, \aap, 614, L3, \dodoi{10.1051/0004-6361/201832934}

\bibitem[{{Gustafsson} {et~al.}(2008){Gustafsson}, {Edvardsson}, {Eriksson}, {J{\o}rgensen}, {Nordlund}, \& {Plez}}]{2008A&A...486..951G}
{Gustafsson}, B., {Edvardsson}, B., {Eriksson}, K., {et~al.} 2008, \aap, 486, 951, \dodoi{10.1051/0004-6361:200809724}

\bibitem[{{Holmbeck} {et~al.}(2023){Holmbeck}, {Sprouse}, \& {Mumpower}}]{Holmbeck2023}
{Holmbeck}, E.~M., {Sprouse}, T.~M., \& {Mumpower}, M.~R. 2023, European Physical Journal A, 59, 28, \dodoi{10.1140/epja/s10050-023-00927-7}

\bibitem[{{Hotokezaka} {et~al.}(2018){Hotokezaka}, {Beniamini}, \& {Piran}}]{2018IJMPD..2742005H}
{Hotokezaka}, K., {Beniamini}, P., \& {Piran}, T. 2018, International Journal of Modern Physics D, 27, 1842005, \dodoi{10.1142/S0218271818420051}

\bibitem[{{Howard} {et~al.}(2012){Howard}, {Marcy}, {Bryson}, {Jenkins}, {Rowe}, {Batalha}, {Borucki}, {Koch}, {Dunham}, {Gautier}, {Van Cleve}, {Cochran}, {Latham}, {Lissauer}, {Torres}, {Brown}, {Gilliland}, {Buchhave}, {Caldwell}, {Christensen-Dalsgaard}, {Ciardi}, {Fressin}, {Haas}, {Howell}, {Kjeldsen}, {Seager}, {Rogers}, {Sasselov}, {Steffen}, {Basri}, {Charbonneau}, {Christiansen}, {Clarke}, {Dupree}, {Fabrycky}, {Fischer}, {Ford}, {Fortney}, {Tarter}, {Girouard}, {Holman}, {Johnson}, {Klaus}, {Machalek}, {Moorhead}, {Morehead}, {Ragozzine}, {Tenenbaum}, {Twicken}, {Quinn}, {Isaacson}, {Shporer}, {Lucas}, {Walkowicz}, {Welsh}, {Boss}, {Devore}, {Gould}, {Smith}, {Morris}, {Prsa}, {Morton}, {Still}, {Thompson}, {Mullally}, {Endl}, \& {MacQueen}}]{2012ApJS..201...15H}
{Howard}, A.~W., {Marcy}, G.~W., {Bryson}, S.~T., {et~al.} 2012, \apjs, 201, 15, \dodoi{10.1088/0067-0049/201/2/15}

\bibitem[{{Issa} {et~al.}(2024){Issa}, {Gottlieb}, {Metzger}, {Jacquemin-Ide}, {Liska}, {Foucart}, {Halevi}, \& {Tchekhovskoy}}]{2024arXiv241002852I}
{Issa}, D., {Gottlieb}, O., {Metzger}, B., {et~al.} 2024, arXiv e-prints, arXiv:2410.02852, \dodoi{10.48550/arXiv.2410.02852}

\bibitem[{{Jellinek} \& {Jackson}(2015)}]{Jellinek2015}
{Jellinek}, A.~M., \& {Jackson}, M.~G. 2015, Nature Geoscience, 8, 587, \dodoi{10.1038/ngeo2488}

\bibitem[{{Kobayashi} {et~al.}(2006){Kobayashi}, {Umeda}, {Nomoto}, {Tominaga}, \& {Ohkubo}}]{2006ApJ...653.1145K}
{Kobayashi}, C., {Umeda}, H., {Nomoto}, K., {Tominaga}, N., \& {Ohkubo}, T. 2006, \apj, 653, 1145, \dodoi{10.1086/508914}

\bibitem[{{Kolborg} {et~al.}(2023){Kolborg}, {Ramirez-Ruiz}, {Martizzi}, {Macias}, \& {Soares-Furtado}}]{Kolborg2023}
{Kolborg}, A.~N., {Ramirez-Ruiz}, E., {Martizzi}, D., {Macias}, P., \& {Soares-Furtado}, M. 2023, \apj, 949, 100, \dodoi{10.3847/1538-4357/acca80}

\bibitem[{{Labrosse}(2014)}]{Labrosse2014}
{Labrosse}, S. 2014, Comptes Rendus Geoscience, 346, 119, \dodoi{10.1016/j.crte.2014.04.005}

\bibitem[{{Labrosse}(2015)}]{Labrosse2015}
---. 2015, Physics of the Earth and Planetary Interiors, 247, 36, \dodoi{10.1016/j.pepi.2015.02.002}

\bibitem[{{Lattimer} \& {Schramm}(1974)}]{1974ApJ...192L.145L}
{Lattimer}, J.~M., \& {Schramm}, D.~N. 1974, \apjl, 192, L145, \dodoi{10.1086/181612}

\bibitem[{{Lichtenegger} {et~al.}(2010){Lichtenegger}, {Lammer}, {Grie{\ss}meier}, {Kulikov}, {von Paris}, {Hausleitner}, {Krauss}, \& {Rauer}}]{Lichtenegger2010}
{Lichtenegger}, H.~I.~M., {Lammer}, H., {Grie{\ss}meier}, J.~M., {et~al.} 2010, \icarus, 210, 1, \dodoi{10.1016/j.icarus.2010.06.042}

\bibitem[{{Lundin} {et~al.}(2007){Lundin}, {Lammer}, \& {Ribas}}]{Ludin2007}
{Lundin}, R., {Lammer}, H., \& {Ribas}, I. 2007, \ssr, 129, 245, \dodoi{10.1007/s11214-007-9176-4}

\bibitem[{{Macias} \& {Ramirez-Ruiz}(2018)}]{Macias2018}
{Macias}, P., \& {Ramirez-Ruiz}, E. 2018, \apj, 860, 89, \dodoi{10.3847/1538-4357/aac3e0}

\bibitem[{{Macias} \& {Ramirez-Ruiz}(2019)}]{Macias2019}
---. 2019, \apjl, 877, L24, \dodoi{10.3847/2041-8213/ab2049}

\bibitem[{{Maoz} \& {Nakar}(2024)}]{2024arXiv240608630M}
{Maoz}, D., \& {Nakar}, E. 2024, arXiv e-prints, arXiv:2406.08630, \dodoi{10.48550/arXiv.2406.08630}

\bibitem[{{Mashonkina} \& {Gehren}(2001)}]{2001A&A...376..232M}
{Mashonkina}, L., \& {Gehren}, T. 2001, \aap, 376, 232, \dodoi{10.1051/0004-6361:20010965}

\bibitem[{{Mello} \& {Fria{\c{c}}a}(2023)}]{Mello2023}
{Mello}, F. d.~S., \& {Fria{\c{c}}a}, A. C.~S. 2023, International Journal of Astrobiology, 22, 272, \dodoi{10.1017/S1473550423000083}

\bibitem[{{Mishenina} {et~al.}(2022){Mishenina}, {Pignatari}, {Gorbaneva}, {C{\^o}t{\'e}}, {Yag{\"u}e L{\'o}pez}, {Thielemann}, \& {Soubiran}}]{2022MNRAS.516.3786M}
{Mishenina}, T., {Pignatari}, M., {Gorbaneva}, T., {et~al.} 2022, \mnras, 516, 3786, \dodoi{10.1093/mnras/stac2361}

\bibitem[{{Mordasini} {et~al.}(2012){Mordasini}, {Alibert}, {Georgy}, {Dittkrist}, {Klahr}, \& {Henning}}]{2012A&A...547A.112M}
{Mordasini}, C., {Alibert}, Y., {Georgy}, C., {et~al.} 2012, \aap, 547, A112, \dodoi{10.1051/0004-6361/201118464}

\bibitem[{{M{\"o}sta} {et~al.}(2018){M{\"o}sta}, {Roberts}, {Halevi}, {Ott}, {Lippuner}, {Haas}, \& {Schnetter}}]{2018ApJ...864..171M}
{M{\"o}sta}, P., {Roberts}, L.~F., {Halevi}, G., {et~al.} 2018, \apj, 864, 171, \dodoi{10.3847/1538-4357/aad6ec}

\bibitem[{{M{\"u}ller} {et~al.}(2024){M{\"u}ller}, {Baron}, {Helled}, {Bouchy}, \& {Parc}}]{2024A&A...686A.296M}
{M{\"u}ller}, S., {Baron}, J., {Helled}, R., {Bouchy}, F., \& {Parc}, L. 2024, \aap, 686, A296, \dodoi{10.1051/0004-6361/202348690}

\bibitem[{{Naiman} {et~al.}(2018){Naiman}, {Pillepich}, {Springel}, {Ramirez-Ruiz}, {Torrey}, {Vogelsberger}, {Pakmor}, {Nelson}, {Marinacci}, {Hernquist}, {Weinberger}, \& {Genel}}]{Naiman2018}
{Naiman}, J.~P., {Pillepich}, A., {Springel}, V., {et~al.} 2018, \mnras, 477, 1206, \dodoi{10.1093/mnras/sty618}

\bibitem[{{Nichols} {et~al.}(2024){Nichols}, {Weiss}, {Eyster}, {Martin}, {Maloof}, {Kelly}, {Zawaski}, {Mojzsis}, {Watson}, \& {Cherniak}}]{Nichols2024}
{Nichols}, C. I.~O., {Weiss}, B.~P., {Eyster}, A., {et~al.} 2024, Journal of Geophysical Research (Solid Earth), 129, e2023JB027706, \dodoi{10.1029/2023JB02770610.31223/x5sx0v}

\bibitem[{Nimmo {et~al.}(2020)Nimmo, Primack, Faber, Ramirez-Ruiz, \& Safarzadeh}]{Nimmo2020}
Nimmo, F., Primack, J., Faber, S.~M., Ramirez-Ruiz, E., \& Safarzadeh, M. 2020, The Astrophysical Journal Letters, 903, L37, \dodoi{10.3847/2041-8213/abc251}

\bibitem[{Ochsenbein(1996)}]{10.26093/cds/vizier}
Ochsenbein, F. 1996, The VizieR database of astronomical catalogues,  CDS, Centre de DonnÃ©es astronomiques de Strasbourg, \dodoi{10.26093/CDS/VIZIER}

\bibitem[{{Ochsenbein} {et~al.}(2000){Ochsenbein}, {Bauer}, \& {Marcout}}]{vizier2000}
{Ochsenbein}, F., {Bauer}, P., \& {Marcout}, J. 2000, \aaps, 143, 23, \dodoi{10.1051/aas:2000169}

\bibitem[{{O'Neill} {et~al.}(2020){O'Neill}, {Lowman}, \& {Wasiliev}}]{O'Neill2020}
{O'Neill}, C., {Lowman}, J., \& {Wasiliev}, J. 2020, \icarus, 352, 114025, \dodoi{10.1016/j.icarus.2020.114025}

\bibitem[{{Otegi} {et~al.}(2020){Otegi}, {Bouchy}, \& {Helled}}]{2020A&A...634A..43O}
{Otegi}, J.~F., {Bouchy}, F., \& {Helled}, R. 2020, \aap, 634, A43, \dodoi{10.1051/0004-6361/201936482}

\bibitem[{Prochaska {et~al.}(2000)Prochaska, Naumov, Carney, McWilliam, \& Wolfe}]{Prochaska_2000}
Prochaska, J.~X., Naumov, S.~O., Carney, B.~W., McWilliam, A., \& Wolfe, A.~M. 2000, The Astronomical Journal, 120, 2513, \dodoi{10.1086/316818}

\bibitem[{{Quick} {et~al.}(2020){Quick}, {Roberge}, {Mlinar}, \& {Hedman}}]{Quick2020}
{Quick}, L.~C., {Roberge}, A., {Mlinar}, A.~B., \& {Hedman}, M.~M. 2020, \pasp, 132, 084402, \dodoi{10.1088/1538-3873/ab9504}

\bibitem[{{Reddy} {et~al.}(2006){Reddy}, {Lambert}, \& {Allende Prieto}}]{2006MNRAS.367.1329R}
{Reddy}, B.~E., {Lambert}, D.~L., \& {Allende Prieto}, C. 2006, \mnras, 367, 1329, \dodoi{10.1111/j.1365-2966.2006.10148.x}

\bibitem[{{Seager} {et~al.}(2007){Seager}, {Kuchner}, {Hier-Majumder}, \& {Militzer}}]{2007ApJ...669.1279S}
{Seager}, S., {Kuchner}, M., {Hier-Majumder}, C.~A., \& {Militzer}, B. 2007, \apj, 669, 1279, \dodoi{10.1086/521346}

\bibitem[{{Shen} {et~al.}(2015){Shen}, {Cooke}, {Ramirez-Ruiz}, {Madau}, {Mayer}, \& {Guedes}}]{2015ApJ...807..115S}
{Shen}, S., {Cooke}, R.~J., {Ramirez-Ruiz}, E., {et~al.} 2015, \apj, 807, 115, \dodoi{10.1088/0004-637X/807/2/115}

\bibitem[{{Siegel}(2019)}]{2019EPJA...55..203S}
{Siegel}, D.~M. 2019, European Physical Journal A, 55, 203, \dodoi{10.1140/epja/i2019-12888-9}

\bibitem[{{Simonetti} {et~al.}(2019){Simonetti}, {Matteucci}, {Greggio}, \& {Cescutti}}]{2019MNRAS.486.2896S}
{Simonetti}, P., {Matteucci}, F., {Greggio}, L., \& {Cescutti}, G. 2019, \mnras, 486, 2896, \dodoi{10.1093/mnras/stz991}

\bibitem[{{Sneden} {et~al.}(2008){Sneden}, {Cowan}, \& {Gallino}}]{Sneden2008}
{Sneden}, C., {Cowan}, J.~J., \& {Gallino}, R. 2008, \araa, 46, 241, \dodoi{10.1146/annurev.astro.46.060407.145207}

\bibitem[{{Sneden}(1973)}]{Sneden1973}
{Sneden}, C.~A. 1973, PhD Thesis, The University of Texas at Austin

\bibitem[{{Spina} {et~al.}(2018){Spina}, {Mel{\'e}ndez}, {Karakas}, {dos Santos}, {Bedell}, {Asplund}, {Ram{\'\i}rez}, {Yong}, {Alves-Brito}, {Bean}, \& {Dreizler}}]{Spina2018}
{Spina}, L., {Mel{\'e}ndez}, J., {Karakas}, A.~I., {et~al.} 2018, \mnras, 474, 2580, \dodoi{10.1093/mnras/stx2938}

\bibitem[{{Tody}(1986)}]{IRAF1}
{Tody}, D. 1986, in Society of Photo-Optical Instrumentation Engineers (SPIE) Conference Series, Vol. 627, Instrumentation in astronomy VI, ed. D.~L. {Crawford}, 733, \dodoi{10.1117/12.968154}

\bibitem[{{Tody}(1993)}]{IRAF2}
{Tody}, D. 1993, in Astronomical Society of the Pacific Conference Series, Vol.~52, Astronomical Data Analysis Software and Systems II, ed. R.~J. {Hanisch}, R.~J.~V. {Brissenden}, \& J.~{Barnes}, 173

\bibitem[{{Toonen} {et~al.}(2012){Toonen}, {Nelemans}, \& {Portegies Zwart}}]{2012A&A...546A..70T}
{Toonen}, S., {Nelemans}, G., \& {Portegies Zwart}, S. 2012, \aap, 546, A70, \dodoi{10.1051/0004-6361/201218966}

\bibitem[{{Unterborn} {et~al.}(2015){Unterborn}, {Johnson}, \& {Panero}}]{Unterborn2015}
{Unterborn}, C.~T., {Johnson}, J.~A., \& {Panero}, W.~R. 2015, \apj, 806, 139, \dodoi{10.1088/0004-637X/806/1/139}

\bibitem[{{Valenti} \& {Fischer}(2005)}]{2005ApJS..159..141V}
{Valenti}, J.~A., \& {Fischer}, D.~A. 2005, \apjs, 159, 141, \dodoi{10.1086/430500}

\bibitem[{{Valenti} \& {Piskunov}(1996)}]{1996A&AS..118..595V}
{Valenti}, J.~A., \& {Piskunov}, N. 1996, \aaps, 118, 595

\bibitem[{{Wang} {et~al.}(2020){Wang}, {Morel}, {Quanz}, \& {Mojzsis}}]{Wang2020}
{Wang}, H.~S., {Morel}, T., {Quanz}, S.~P., \& {Mojzsis}, S.~J. 2020, \aap, 644, A19, \dodoi{10.1051/0004-6361/202038386}

\bibitem[{{Wenger} {et~al.}(2000){Wenger}, {Ochsenbein}, {Egret}, {Dubois}, {Bonnarel}, {Borde}, {Genova}, {Jasniewicz}, {Lalo{\"e}}, {Lesteven}, \& {Monier}}]{2000A&AS..143....9W}
{Wenger}, M., {Ochsenbein}, F., {Egret}, D., {et~al.} 2000, \aaps, 143, 9, \dodoi{10.1051/aas:2000332}

\bibitem[{{Wienbruch} \& {Spohn}(1995)}]{Wienbruch1995}
{Wienbruch}, U., \& {Spohn}, T. 1995, \planss, 43, 1045, \dodoi{10.1016/0032-0633(95)00039-8}

\bibitem[{{Winckler} {et~al.}(2006){Winckler}, {Dababneh}, {Heil}, {K{\"a}ppeler}, {Gallino}, \& {Pignatari}}]{2006ApJ...647..685W}
{Winckler}, N., {Dababneh}, S., {Heil}, M., {et~al.} 2006, \apj, 647, 685, \dodoi{10.1086/505026}

\bibitem[{{Wiseman} {et~al.}(2023){Wiseman}, {Sullivan}, {Smith}, \& {Popovic}}]{2023MNRAS.520.6214W}
{Wiseman}, P., {Sullivan}, M., {Smith}, M., \& {Popovic}, B. 2023, \mnras, 520, 6214, \dodoi{10.1093/mnras/stad488}

\bibitem[{{Woosley} \& {Weaver}(1995)}]{1995ApJS..101..181W}
{Woosley}, S.~E., \& {Weaver}, T.~A. 1995, \apjs, 101, 181, \dodoi{10.1086/192237}

\bibitem[{{Zha} {et~al.}(2024){Zha}, {M{\"u}ller}, \& {Powell}}]{2024ApJ...969..141Z}
{Zha}, S., {M{\"u}ller}, B., \& {Powell}, J. 2024, \apj, 969, 141, \dodoi{10.3847/1538-4357/ad4ae7}

\bibitem[{{Zhang} \& {Rogers}(2022)}]{Zhang2022}
{Zhang}, J., \& {Rogers}, L.~A. 2022, \apj, 938, 131, \dodoi{10.3847/1538-4357/ac8e65}

\end{thebibliography}
\bibliographystyle{aasjournal}

\end{document}